\documentclass{article}
\usepackage{spconf,amsmath,graphicx,hyperref}
\usepackage{cite}
\usepackage{amsmath,amssymb,amsfonts}
\usepackage{algorithmic}
\usepackage{graphicx}
\usepackage{textcomp}
\usepackage{xcolor}
\usepackage{graphicx}
\usepackage{textcomp}
\usepackage{xcolor}
\usepackage{algorithm2e}
\RestyleAlgo{ruled}
\usepackage{lineno}
\usepackage{array}
\usepackage{multirow}
\usepackage{slashbox}
\usepackage{balance, flushend}
\usepackage{tikz}
\usetikzlibrary{calc}
\usepackage{booktabs}
\usepackage{arydshln}
\usetikzlibrary{arrows.meta}
\usetikzlibrary{shapes.geometric}
\usepackage{amssymb}% http://ctan.org/pkg/pifont

\usepackage{fancyhdr}
\usepackage{textcomp}
\usepackage[absolute]{textpos}
\newcommand{\copyrightstatement}{
	\begin{textblock}{15}(0.4,0.2)
	\noindent
	\textblockcolour{white}
	\copyright 2026 IEEE. Published in 2026 International Conference on Image Processing (ICIP), scheduled for 13-17 September 2026 in Tampere, Finnland. Personal use of this material is permitted. However, permission to reprint/republish this material for advertising or promotional purposes or for creating new collective works for resale or redistribution to servers or lists, or to reuse any copyrighted component of this work in other works, must be obtained from the IEEE. DOI: \url{10.1109/ICIP61757.2026.11630382}
	\end{textblock}
}

\title{HIERARCHICAL FILTER BAND SELECTION FOR MULTISPECTRAL OBJECT CLASSIFICATION}
\name{Katja Kossira, Jürgen Seiler, and André Kaup \thanks{The authors gratefully acknowledge that this work has been supported by the Bayerische Forschungsstiftung (BFS, Bavarian Research Foundation) under project number AZ-1547-22.}}
\address{Multimedia Communications and Signal Processing\\Friedrich-Alexander-Universität Erlangen-Nürnberg\\
	Cauerstr. 7, 91058, Erlangen, Germany\\
	\{katja.kossira, juergen.seiler, andre.kaup\} @fau.de}

\begin{document}
%\ninept
%
\maketitle
\begin{abstract}
\vspace{-0.2cm}
Multispectral camera arrays capture image data in various spectral bands, enabling image acquisition beyond human perception. These systems are widely used in medical, agricultural, environmental, and remote sensing applications. However, not all recorded bands are needed for classification tasks, thus reducing them can lower hardware complexity and cost. The conditional filter band selection algorithm addresses this by selecting low-noise, non-redundant bands to minimize the number of filters and cameras. This paper improves the approach by introducing a second classification stage that estimates object material in addition to the object label. This information is merged by a decision-tree based band selection strategy. The proposed method achieves a 28.9\% relative reduction in classification error on the SMM50 dataset compared to the state-of-the-art. Moreover, for the same classification accuracy, the required number of cameras is reduced from 7 to 4, demonstrating that the proposed approach improves performance while significantly lowering hardware requirements.

\end{abstract}


\vspace{-0.4cm}
\section{Introduction}
\vspace{-0.2cm}
\label{sec:Introduction}
In recent decades, multispectral imaging (MSI) has gained importance across various fields. This interest is based on the fact that different materials absorb and reflect light at specific wavelengths, enabling applications such as palm vein detection \cite{VeinDetection}, heart rate estimation \cite{HeartRate}, food processing and agricultural monitoring \cite{CropMonitoring}, sorting materials for recycling \cite{Recycling}, or remote sensing for environmental monitoring \cite{Mosquito}. 

\begin{figure}
	\centering
	\begin{tikzpicture}
		\node(CAMSI)[]{\includegraphics[width=0.2\textwidth]{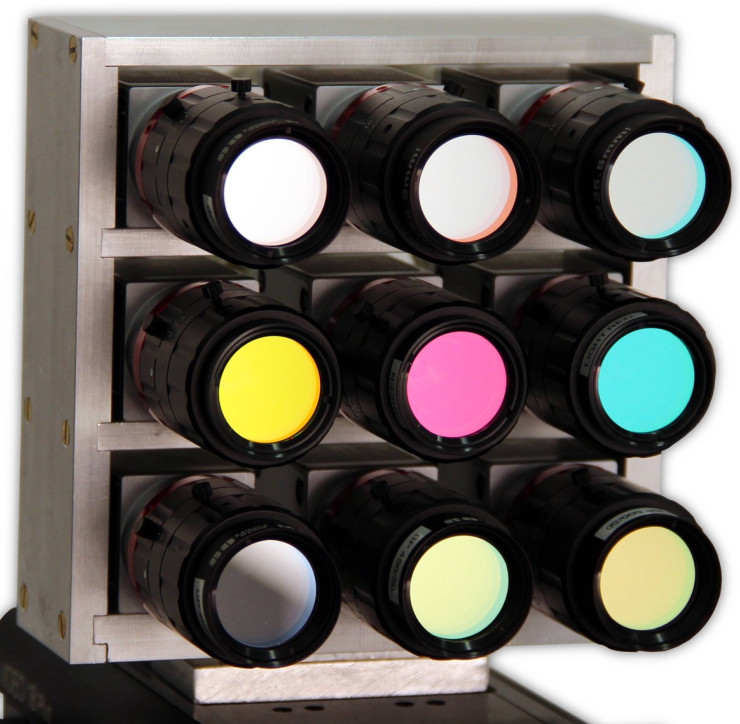}};
	\end{tikzpicture}
	\vspace{-0.5cm}
	\caption{Example of a multispectral camera array \cite{FBS} with $n=9$ channels.}
	\label{fig:CAMSI}
	\vspace{-0.5cm}
\end{figure}

MSI setups can be realized in several ways. One method involves multiplexed illumination, where multiple filtered LED subsets are activated during image acquisition \cite{Multiplexed}. Another approach uses a filter wheel with $n$ filters in front of a single camera \cite{FilterWheel}. As the wheel rotates, $n$ images of the same scene are captured sequentially. Since this method records images one after another, a stable environment is required and moving video capture is not possible. Monno et al. extended the classical Bayer pattern \cite{RGBBayer} used for RGB imaging to multispectral filter arrays (MSFAs) \cite{MSFA}, where a periodic filter pattern is applied across a grayscale sensor. However, demosaicing similar to RGB Bayer interpolation is required, reducing the resolution. Camera arrays for multispectral or hyperspectral imaging \cite{FBS}, as exemplary shown in Figure \ref{fig:CAMSI}, are multi-camera, multi-filter configurations, which allow simultaneous image and video recording. 
Further, each of the monochromatic cameras is equipped with a filter mounted in front of its lens, which can be exchanged to suit different applications and spectral ranges, resulting in a large number of possible combinations. 

\begin{figure}
	\centering
	\begin{tikzpicture}
		\node(root)[fill=red!30, rounded corners=6pt, inner sep=5pt]{Root node};
		\node(classl)[yshift=-0.5cm, xshift=0.9cm]{\color{gray}Class label};
		\node(decision1)[below of=root, fill=green!20, rounded corners=6pt, inner sep=5pt, xshift=-1.8cm, yshift=-0.4cm]{Decision node};
		\node(objl1)[below of=decision1, yshift=0.5cm, xshift=0.9cm]{\color{gray}Object label};
		\node(decision2)[below of=root, fill=green!20, rounded corners=6pt, inner sep=5pt, xshift=1.8cm, yshift=-0.4cm]{Decision node};
		\node(objl2)[below of=decision2, yshift=0.5cm, xshift=0.9cm]{\color{gray}Object label};
		\node(leaf1)[below of=decision1, fill=blue!20, rounded corners=8pt, inner sep=5pt, xshift=-0.9cm, yshift=-0.4cm]{Leaf node};
		\node(leaf2)[below of=decision1, fill=blue!20, rounded corners=8pt, inner sep=5pt, xshift=0.9cm, yshift=-0.4cm]{Leaf node};
		\node(leaf3)[below of=decision2, fill=blue!20, rounded corners=8pt, inner sep=5pt, xshift=0.9cm, yshift=-0.4cm]{Leaf node};
		\node(leaf4)[below of=decision2, fill=blue!20, rounded corners=8pt, inner sep=5pt, xshift=-0.9cm, yshift=-0.4cm]{Leaf node};
		\draw[->] (root) -- ++(0,-0.7) -| (decision1.north);
		\draw[->] (root) -- ++(0,-0.7) -| (decision2.north); 
		\draw[->] (decision1) -- ++(0,-0.7) -| (leaf1.north);
		\draw[->] (decision1) -- ++(0,-0.7) -| (leaf2.north);
		\draw[->] (decision2) -- ++(0,-0.7) -| (leaf3.north);
		\draw[->] (decision2) -- ++(0,-0.7) -| (leaf4.north);
		\node(obj)[right of=root, xshift=3cm, yshift=-0.3cm]{\color{darkgray}All objects};
		\node(class)[below of=obj, yshift=-0.2cm, xshift=0.2cm]{\color{darkgray}Classes};
		\node(aobj)[below of=class, yshift=-0.3cm]{\color{darkgray}Objects};
		\draw[dashed, color=darkgray] (-3.5,-0.8) -- (4.6,-0.8); 
		\draw[dashed, color=darkgray] (-3.5,-2.2) -- (4.6,-2.2);
	\end{tikzpicture}
	\vspace{-0.2cm}
	\caption{Principle of the employed decision tree logic \cite{DecisionTree}.}
	\label{fig:DecisionTree}
	\vspace{-0.4cm}
\end{figure}
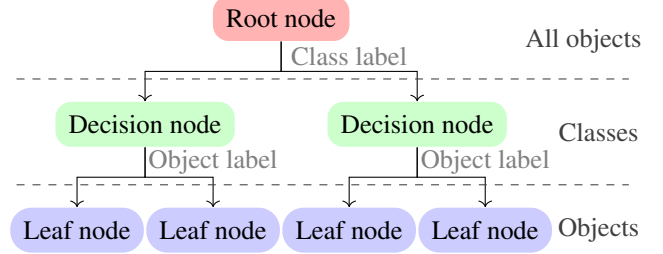

To evaluate all possible filter combinations for such camera setups, a variety of techniques have been proposed. Among them, the fast binary search (FBS) \cite{FBS} and conditional filter band selection (CFBS) \cite{CFBS} algorithms represent two of the most recent approaches. FBS evaluates all spectral bands in order to maximize the spectral angle, which makes it computationally expensive and potentially ineffective. CFBS addresses this limitation by additionally incorporating noise characteristics into the selection process, thereby improving robustness. However, CFBS does not exploit all information that is actually available in the data. In many application domains, such as recycling, objects are not only associated with object labels but also with material classes (e.g. wood, plastic) \cite{ScienceTagesschau}. These material labels can provide valuable prior knowledge that has so far not been considered in existing filter band selection methods. This becomes particularly relevant as the availability of new datasets for classification tasks is increasing. While dual-labeled datasets (e.g. object type and material) are still relatively rare, their emergence reflects the growing importance of multimodal classification settings. Integrating such complementary information can help avoid misclassifications in challenging scenarios. To overcome this limitation, a hierarchical filter band selection (HFBS) algorithm is proposed. In contrast to previous methods, HFBS utilizes a bi-modal input stage processed through a hierarchical fusion rule, where a second classification step determines the material type of the object. This information is then combined with the object label through a decision tree logic \cite{DecisionTree}, as illustrated in Fig. \ref{fig:DecisionTree}. The root node represents all objects, decision nodes correspond to material classes, and leaf nodes yield the final object classification. As the filter selection is guided by a hierarchical label structure rather than task-specific classifiers, the proposed approach can be transferred to different sensing setups and application domains with minimal adaption of the training data.

The paper is structured as follows. Section \ref{sec:related work} reviews the current state of the art. Section \ref{sec:proposed method} introduces CFBS for a minimal filter selection and the novel HFBS approach for improved hierarchical filter band selection. The results are presented and verified in Section \ref{sec:verification}, with detailed comparison to conventional solutions. Summary and conclusion are given in Section \ref{sec:Conclusion}.
\vspace{-0.4cm}
\section{Related Work}
\vspace{-0.3cm}
\label{sec:related work}
In the past, various methods have been developed to select suitable filters for multispectral object classification. Heuristic methods like uniform search \cite{UniformSearch} distribute filters evenly across the spectrum to ensure coverage. However, due to the limited number of filters in camera arrays, wide-band filters are often necessary to avoid spectral gaps. Hardenberg et al. \cite{Hardenberg} propose maximizing orthogonality in the characteristic reflectance vector space (MOCR) representative of the application area for the system. The filters are chosen sequentially and aim to capture areas most spectra have in common. Thus, the regions that make the spectra distinguishable are not selected. Beyond the sequential methods, Kumar et al. \cite{RemoteSensing} proposed a framework that integrates spatial-spectral features, while Feng et al. \cite{RemoteSensing2} introduced an unsupervised dual graph autoencoder guided by diversity learning to select representative bands. In contrast, the brute force evaluates all possible filter combinations to find the optimal subset. As every possible combination of filters from the available pool is considered and tested, this solution is computationally expensive, especially for large filter sets. For example, when selecting 9 out of 150 available filters, more than $8 \times 10^{12}$ combinations are possible and have to be evaluated. To reduce this effort, Sippel et al. \cite{FBS} developed a fast binary search (FBS) algorithm, which ensures optimal selection of $N$ out of $K$ filters with significantly lower computational cost by considering the spectral angle between filter bands. 

Since the number of filters to chose has to be previously determined in FBS, the conditional filter band selection (CFBS) was introduced in \cite{CFBS}, which is based on FBS, but allows to select a minimum number of filters required, while still guaranteeing a good multispectral object classification. Furthermore, this approach enables noise consideration, which improves classification by favoring less noisy adjacent bands over distorted ones. As a result, reliability and overall performance are increased.

\vspace{-0.3cm}
\section{Hierarchical Filter Band Selection}
\begin{figure}[t]
	\centering
	\begin{tikzpicture}[auto]
		\node(filterset) [fill=white, minimum height=0.7cm, text width=1.5cm, align=center, xshift=-1.4cm] {Filterset,\\object \\ \textcolor{gray}{White oak}};
		\node (CFBS) [right of = filterset, draw, fill=white,minimum height=0.9cm, align=center, xshift = 0.7cm] {CFBS};
		\draw[->](filterset) -- (CFBS);
		\node(objclass)[align=center, draw, fill=white, right of=CFBS, xshift=0.8cm, yshift=-0.7cm]{Object\\classification};
		\node(classclass)[align=center, draw, fill=white, above of=objclass, yshift=0.4cm]{Material\\classification};
		\draw[->](CFBS) |- (classclass);
		\draw[->](CFBS) |- (objclass);
		\node(sum)[draw, circle, right of=CFBS, xshift=2.8cm]{$+$};
		\node(label)[align=center, right of=sum, xshift=1.05cm]{Object label\\\color{gray}White oak};
		\draw[->](classclass) -|node[above]{\textcolor{gray}{Wood}} (sum);
		\draw[->](objclass) -|node[below,align=center]{\textcolor{gray}{White oak;}\\\color{gray}polyethylene} (sum);
		\draw[->] (sum)--(label);
		\draw[blue, dashed, fill=blue!20, opacity=0.5] ($(CFBS.north west) + (-0.2,0.9)$) rectangle ($(classclass.south east) + (1.9,-1.9)$);
		\node (CFBS) [right of = filterset, draw, fill=white,minimum height=0.9cm, align=center, xshift = 0.7cm] {CFBS};
		\draw[->](filterset) -- (CFBS);
		\node(objclass)[align=center, draw, fill=white, right of=CFBS, xshift=0.8cm, yshift=-0.7cm]{Object\\classification};
		\node(classclass)[align=center, draw, fill=white, above of=objclass, yshift=0.4cm]{Material\\classification};
		\draw[->](CFBS) |- (classclass);
		\draw[->](CFBS) |- (objclass);
		\node(sum)[draw, circle, right of=CFBS, xshift=2.8cm]{$+$};
		\node(label)[align=center, right of=sum, xshift=1.05cm]{Object label\\\color{gray}White oak};
		\draw[->](classclass) -|node[above]{\textcolor{gray}{Wood}} (sum);
		\draw[->](objclass) -|node[below,align=center]{\textcolor{gray}{White oak;}\\\color{gray}polyethylene} (sum);
		\draw[->] (sum)--(label);
		\node [below of = objclass, xshift=0.4cm, yshift=-0.2cm, blue] {HFBS};
	\end{tikzpicture}
	\vspace{-0.9cm}
	\caption{Pipeline and example of the proposed HFBS algorithm.}
	\label{fig:TS-CFBS principle}
	\vspace{-0.3cm}
\end{figure}
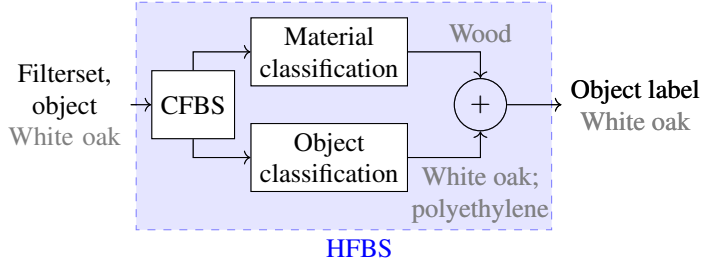

\vspace{-0.3cm}
\label{sec:proposed method}
In \cite{CFBS}, conditional filter band selection (CFBS) was shown to perform well using the spectral measurements of materials (SMM50) databases \textit{scio} and \textit{lumini} \cite{SMM}, achieving a significant reduction in the number of wrongly classified objects (WCO) compared to the FBS \cite{FBS} algorithm. However, the method does not utilize the valuable information provided by the material class of the object. To incorporate this data, a novel hierarchical filter band selection (HFBS) algorithm is proposed, the pipeline of which is shown in Fig. \ref{fig:TS-CFBS principle}. Following CFBS, two parallel classification processes are performed on the same input data. While one process predicts the object label, the other estimates the object material class. The resulting material label is then used to refine the object classification based on the decision tree principle \cite{DecisionTree}, as depicted in Fig. \ref{fig:DecisionTree}.

Given are $K$ spectral bands from a predefined filter set $\mathcal{F}=\{f_1, f_2,...,f_K\}$, where each $f_i$ corresponds to a specific wavelength. The CFBS module first selects an optimal subset of $N$ out of $K$ bands according to a discriminative criterion $\mathcal{J}(\cdot)$, such that
\begin{equation}
	\mathcal{F}^\ast = \underset{S \subset \mathcal{F},\, |S|=N}{\arg\max} \; \mathcal{J}(S) ~~~ .
\end{equation}
CFBS returns the minimal optimal filter subset $\mathcal{F^\ast}$, i.e., the smallest subset that maximizes the spectral angle $\theta$ between adjacent spectral filter bands $\textbf{\textit{f}}_i, \textbf{\textit{f}}_j$
\begin{equation}
	\theta(\textbf{\textit{f}}_{i}, \textbf{\textit{f}}_{j}) = \arccos \left(\frac{\textbf{\textit{f}}_{i}^{T}}{||\textbf{\textit{f}}_{i}||_{2}} \frac{\textbf{\textit{f}}_{j}}{||\textbf{\textit{f}}_{j}||_{2}}\right) ~~~.
\end{equation}

To ensure that the selected bands provide maximal useful information while avoiding redundancy, CFBS applies two complementary evaluations. First, inter-band correlations are assessed to identify and avoid redundant spectral information. Second, the noise level of each band is considered to favor stable and informative measurements. In this scenario, this combination captures the dominant characteristics of the spectral data, and additional statistical criteria were not observed to yield further performance gains. 

\begin{figure}[t!]
	\centering
	\includegraphics[width=0.38\textwidth]{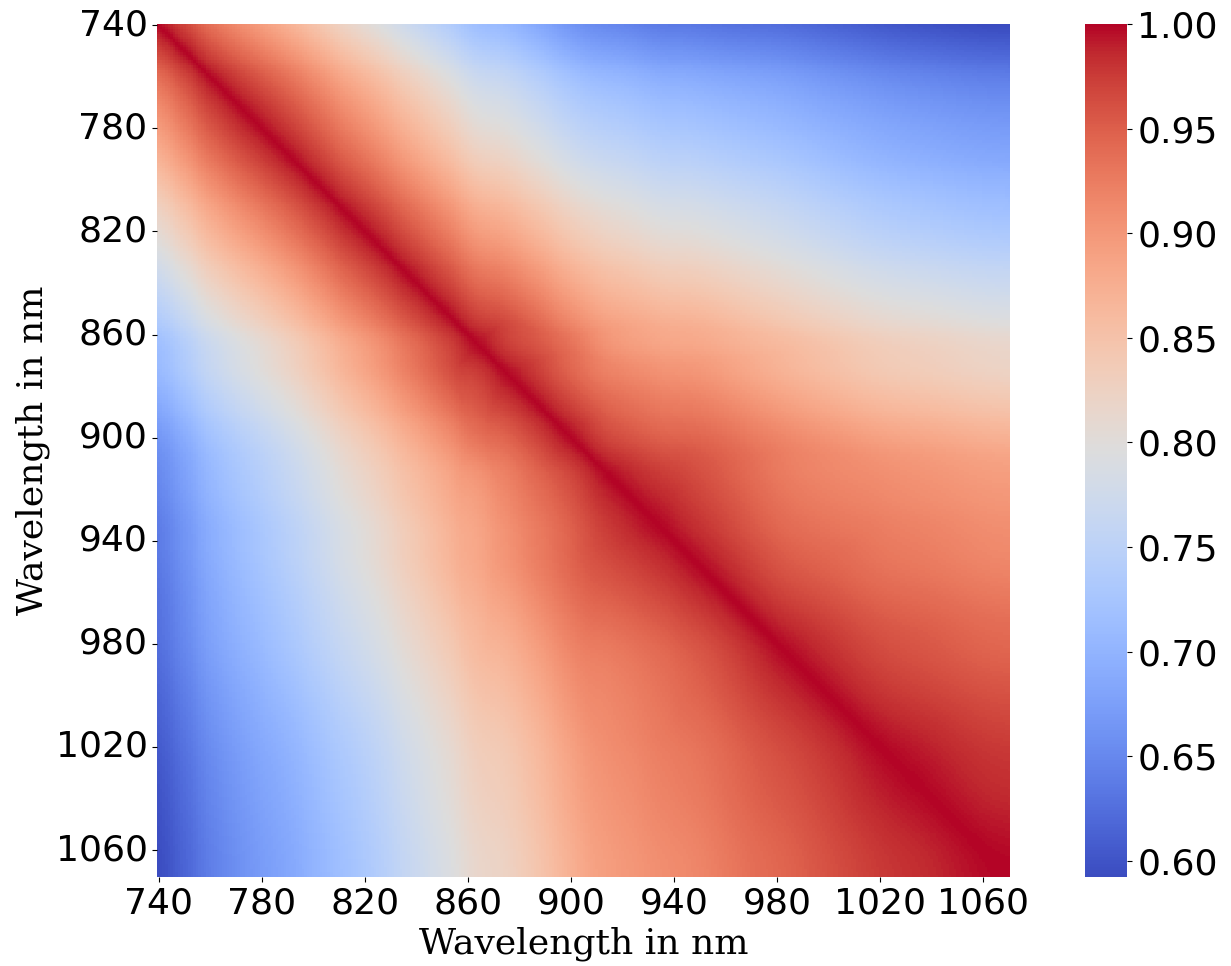}
	\vspace{-0.4cm}
	\caption{Exemplary Kendall correlation matrix of the SMM50 dataset \cite{SMM}. Strong correlations are highlighted in red, weak correlations are highlighted in blue.}
	\label{fig:KendallCorrelationMatrix}
	\vspace{-0.4cm}
\end{figure}

The Kendall correlation \cite{Kendall} is applied to quantify dependencies among the spectral bands. The resulting Kendall correlation matrix of the SMM50 dataset \cite{SMM}, shown in Fig. \ref{fig:KendallCorrelationMatrix}, highlights strong correlations in red and weak ones in blue. The analysis illustrates the high degree of similarity between neighboring spectral bands, motivating a selection strategy in which redundant adjacent wavelengths are replaced by the locally least noisy representative band.

The Signal-to-Noise Ratio (SNR) \cite{SNR} SNR$_{k}$ of each of the $K$ filter signals $f_{k}$ can be estimated using the ratio of mean $\mu_{k}$  and standard deviation $\sigma_{k}$:
\begin{equation}
\vspace{-0.1cm}
	\text{SNR}_{k} = \frac{\mu_{k}}{\sigma_{k}} = \frac{\frac{1}{M} 	\sum_{i = 1}^{M} x_{k}[m,n]}{\sqrt{\frac{1}{M} \sum_{i = 1}^{M} 	(x_{k}[m,n] - \mu_{k})^{2}}} ~~~, 
	\label{eq:SNR}
	\vspace{-0.1cm}
\end{equation}
where $M$ corresponds to the number of pixels in the signal region and $x_{k}[m,n]$ depict the single pixel values of the $k$-th image. This formulation is a practical adaption for imaging data where the signal power is not explicitly known. Bands with an SNR$_{k}$ below a certain SNR-threshold $\Theta$ are excluded from further consideration as noise degrades the classification performance.
Thus, the combination of correlation analysis and noise constraints yields a reduced set of spectrally non-redundant and reliably measurable bands.

\begin{figure}
	\centering
	\includegraphics[width=0.45\textwidth]{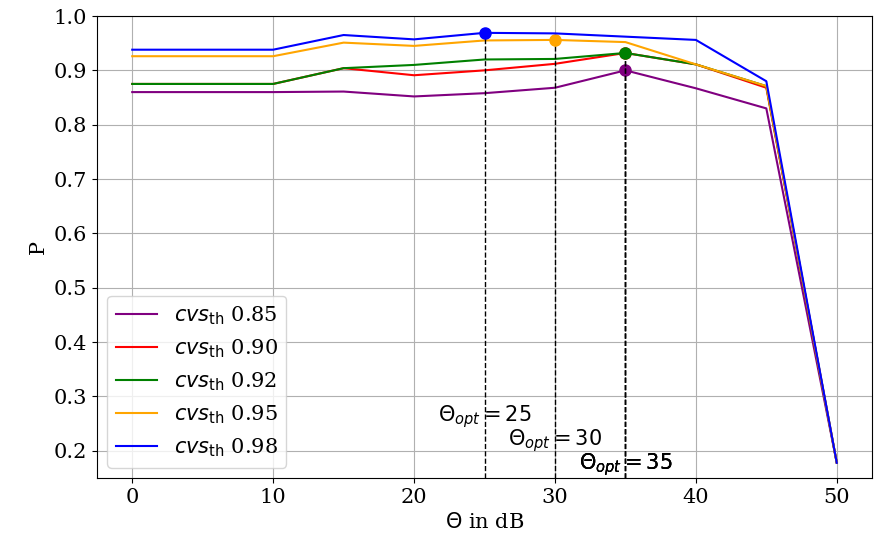}
	\vspace{-0.4cm}
	\caption{Cross-validation score performance \textit{P} for different $cvs_{\text{th}}$ and $\Theta$ combinations.}
	\label{fig:cvsSNRcombinations}
	\vspace{-0.5cm}
\end{figure}

The discriminative criterion $\mathcal{J}(\cdot)$ is evaluated under a cross-validation score threshold $cvs_{\text{th}}$, which specifies the minimum cross-validated classification score that a band combination must achieve to be accepted as sufficiently informative. $\mathcal{J}(\cdot)$ is computed after the hierarchical fusion rule, which means that the quality of a band subset is determined by its ability to support the combined decision of material $y_{\text{mat}}$ and object $y_{\text{obj}}$. Consequently, the early stopping criterion $cvs_{\text{th}}$ is reached when the fused decision meets the performance target, and the selection process can terminate, thereby returning the minimum number of filters required to meet the desired performance. By systematically varying $\Theta$ and $cvs_{\text{th}}$ and exhaustively evaluating all possible $N$ out of $K$ filter band combinations for the datasets, it is quantified how these parameters affect classification performance and hardware requirements. These experiments are summarized in Fig. \ref{fig:cvsSNRcombinations} and Table \ref{tab:ResultsTS-CFBS}. Fig. \ref{fig:cvsSNRcombinations} shows the relationship between the SNR-threshold and the cross-validation score threshold for the SMM50 datasets \cite{SMM}, while Table \ref{tab:ResultsTS-CFBS} lists the corresponding number of wrongly classified objects and the number of filters $n$ required for each parameter combination. 
\begin{table}[t]
	\centering
	\caption{The number of filters $n$ and the number of $\text{WCO}$ achieved using HFBS with different $\Theta$ and $cvs_{\text{th}}$. Results are reported as average across the \textit{lumini} and \textit{scio} sensors.} 
	\begin{tabular}{c|c|cccccc}
		\multicolumn{2}{c}{}& \multicolumn{6}{|c}{$\Theta$} \\ \cline{3-8}
		\multicolumn{2}{r|}{$cvs_{\text{th}}$} & 0 & 10 & 20 & 30 & 40 & 50 \\   
		\hline
		\hline
		\multirow{4}{*}{$n$}& 0.98 & 7 & 7 & 5 & 5 & 5 & 9 \\
		& 0.95 & 6 & 6 & 4 & 4 & 3 & 9 \\
		& 0.92 & 5 & 4 & 3 & 3 & 3 & 9 \\
		& 0.90 & 5 & 4  & 3 & 3 & 3 & 9 \\
		\hline
		\multirow{4}{*}{$\text{WCO}$} & 0.98 & 75 & 73 & 55 & 44 & 53 & 993 \\
		& 0.95 & 89 & 89 & 67 & 53  & 134 & 1289 \\
		& 0.92 & 188 & 188 & 141 & 130 & 134 & 1334 \\
		& 0.90 & 188 & 188 & 160 & 139 & 178 & 1334 \\
	\end{tabular}
	\label{tab:ResultsTS-CFBS}
	\vspace{-0.3cm}
\end{table}

Analysis of these results show that the optimal SNR-threshold $\Theta_{\text{opt}}$ typically lies between 25 and 35 for most $cvs_{\text{th}}$ settings. Very high thresholds such as $\Theta = 50$ remove many higher wavelength bands due to elevated noise levels, leaving only closely spaced bands that are highly correlated. Consequently, more wavelengths must be chosen to compensate for the reduced per-band information. Conversely, very low $\Theta$-values include noisy bands that lower the effective information content and thus reduce classification performance.

The spectral data $I_{\mathcal{F^\ast}} \in \mathbb{R}$ obtained with the reduced filter set $\mathcal{F^\ast}$ is then passed to two classifiers $C_{\text{mat}}$ and $C_{\text{obj}}$, which are operating in parallel. $C_{\text{mat}}$ predicts the material label according to $y_{\text{mat}} = C_{\text{mat}}(\mathcal{I_{F^\ast}})$ and $C_{\text{obj}}$ predicts the object label as $y_{\text{obj}} = C_{\text{obj}}(\mathcal{I_{F^\ast}})$. The outputs are combined through a fusion function $\Phi(\cdot,\cdot)$:
\begin{equation}
	\Phi(y_{\text{mat}}, y_{\text{obj}}) = y_{\text{obj}} \;|\; y_{\text{mat}}~~~,
\end{equation}
which represents a hierarchical decision rule in which object predictions $y_{\text{obj}}$ are evaluated within the context of the predicted material class $y_{\text{mat}}$. E.g., if \text{$y_{\text{mat}}$ = \texttt{Wood}}, \text{$y_{\text{obj1}}$ = \texttt{Polyethylene}} and \text{$y_{\text{obj2}}$ = \texttt{White oak}}, the fused decision corresponds to the final object label \text{$\hat{y}$= \texttt{White oak}}. This hierarchical approach reduces spectral dimensionality, lowers computational load, and improves robustness by making the filter selection process task-aware, while the hierarchical fusion mitigates the risk of misclassification in either branch. Thus, formally, the overall decision process can be expressed as
\begin{equation}
	\hat{y} = \Phi (C_{\text{mat}}(\text{CFBS}(\mathcal{I})), C_{\text{obj}}(\text{CFBS}(\mathcal{I})))~~~,
\end{equation}
where CFBS($\cdot$) denotes the optimized band selection operator.

As HFBS leverages a material-guided decision tree to systematically narrow the set of candidate classifications, it can substantially reduce the number of misclassifications. The algorithm provides a tunable trade-off between classification performance and acquisition cost. By adjusting $cvs_{\text{th}}$ and $\Theta$, the filter selection can be optimized either for maximal classification accuracy or for a minimal number of required cameras and filters, thereby enabling cost-effective solutions tailored to specific applications. 
	
\vspace{-0.3cm}
\section{Performance Evaluation}
\label{sec:verification}
\vspace{-0.3cm}
The hierarchical filter band selection algorithm is evaluated using the publicly available SMM50 datasets \textit{scio} and \textit{lumini} \cite{SMM}, which represent one of the few annotated multispectral datasets available for filter selection and wavelength optimization. The datasets contain five material classes (metal, plastic, wood, paper, fabric), each comprising 10 distinct objects, resulting in 50 independent measurements. Each measurement provides 100 high-resolution spectra sampled at 3 nm intervals in the visible and near-infrared range. The \textit{scio} and \textit{lumini} datasets were acquired using two different spectrometers, capturing the same objects under identical conditions. To show how adaptable HFBS is across setups, we validated the experiments on both spectrometers and the average results across both sensors are reported, what enables a controlled cross-sensor evaluation of the proposed method. Given that the focus is optimal filter and wavelength selection for classification rather than object classification itself, the evaluation emphasizes spectral consistency across sensors instead of dataset scale.
	
The objects are classified using a Gradient Boosting (GB) \cite{GradientBoosting} and Random Forest (RF) \cite{RandomForest} classifier. The former algorithm builds decision trees sequentially, with each tree designed to correct the errors of its predecessors. This approach is powerful but can be less robust to overfitting compared to RF. In contrast, the RF algorithm constructs multiple independent decision trees in parallel, each trained on a random subset of the training data and a random subset of features. The final classification is based on majority voting across all trees, which increases robustness against overfitting. 
	
Both classifiers are implemented with 300 estimators, a learning rate of 0.1, a maximum tree depth of 3 and a minimum sample split of 2. The model performance is evaluated using k-fold cross-validation, ensuring that the train and test splits are stratified to maintain the class distribution across folds. This setup prevents overfitting to a single train-test split and provides a more robust estimate of real-world performance.
	
For the filter simulation, a virtual filter set consisting of 40 bandpass filters is designed, each with a bandwidth of 50 nm, to approximate the characteristics of real camera filters. Normalization parameters obtained from the training step are applied to the recorded multispectral bands. This step is crucial as higher values resulting from a higher filter bandwidth would result in a stronger influence on the spectral angle. The RF and GB classifiers are employed to classify the spectrum of interest, leading to a class label and an object label. Incorporating the material class into the final decision-making step helps to reduce misclassifications.
	
The performance is evaluated on the accuracy, the number of WCO as well as the number of required filters $n$. Based on these metrics, two complementary evaluation scenarios are defined allowing for an ablation study of the trade-off between classification performance and hardware efficiency. The first scenario focuses on minimizing the WCO while maintaining a fixed $n$, prioritizing the classification accuracy. The second scenario aims to reduce the number of cameras required for a specific accuracy level, optimizing the system's efficiency and hardware requirements, such as cameras and filters. 

\vspace{-0.3cm}	
\subsection{Optimizing \textit{n} for fixed WCO}
\vspace{-0.3cm} 
Compared to existing approaches from literature, our HFBS requires significantly fewer cameras to achieve a fixed WCO, as summarized in Table \ref{tab:fixWCOvaryingn}. 
	\begin{table}[t!]
		\centering
		\caption{Required number of cameras $n$ of different approaches from literature for a fixed limit of WCO.}
		\vspace{0.1cm}
		\begin{tabular}{l|ccccc}
			\backslashbox{Method}{WCO} & ~~350~ & ~320~ & ~318~ & ~124~~ & ~~73~~ \\
			\hline
		
			Uniform \cite{UniformSearch} & 9 & n/a & n/a & n/a & n/a\\
			MOCR \cite{Hardenberg} & 9 & 9 & n/a & n/a & n/a \\
			FBS \cite{FBS} & 9 & 9 & 9 & n/a & n/a \\
			CFBS \cite{CFBS} & 4 & 4 & 4 & 7 & n/a \\
			HFBS (ours) & 3 & 3 & 3 & 4 & 7 \\
	
			BF (optimal) & 2 & 2 & 3 & 4 & 6 		
		\end{tabular}
		\label{tab:fixWCOvaryingn}
		\vspace{-0.5cm}
	\end{table}
While uniform search \cite{UniformSearch}, MOCR \cite{Hardenberg}, or FBS \cite{FBS} consistently require the maximum number of $n = 9$ cameras and are not able to reach stricter WCO limits, CFBS \cite{CFBS} reduces the camera count to $n = 4$ for comparable accuracy. HFBS further improves upon this result by achieving the same WCO limits with only \mbox{$n = 3$} cameras, demonstrating a substantially more efficient usage of spectral information. Across all reported WCO limits, HFBS consistently requires fewer cameras than CFBS and all other reference methods.

The brute-force (BF) solution represents the optimal camera selection, as it exhaustively evaluates all possible filter combinations for $n = 1,\dots,9$ and therefore achieves the minimum possible number of cameras. However, as shown in Table \ref{tab:compCost}, this comes at the cost of prohibitively high computational complexity, making BF infeasible for practical applications. Adding even a single additional filter drastically increases the runtime of the BF approach, as the number of combinations to be evaluated grows exponentially with the available filter set $K$ (i.e. $\mathcal{O}(2^K)$). In contrast, HFBS operates very close to the BF optimum and requires at most one additional camera compared to BF, while reducing the computational cost by several orders of magnitude. 
	
\vspace{-0.3cm}
\subsection{Optimizing WCO for fixed \textit{n}}
\vspace{-0.3cm}
For economical applications, it can be advantageous to fix the number of cameras while aiming for the lowest possible WCO, thereby maximizing classification accuracy. In this context, HFBS demonstrates substantially superior performance, achieving far fewer \-misclassifications for a fixed $n$, as shown in Table \ref{tab:fixnvaryingWCO}. For $n = 8$ and $n = 9$, HFBS and CFBS converge to the same solutions as the SNR-threshold eliminates many bands, so that both algorithms ultimately select the same wavelengths with nearly identical spectral information. This redundancy can lead to overfitting and higher misclassifications. 

For smaller values of $n$, specifically \mbox{$n = 3,\dots,7$}, HFBS consistently produces solutions very close to the BF optimum, whereas all reference methods show substantially higher WCO values. This becomes particularly evident at $n = 7$, where HFBS achieves a WCO of only 73 compared to 111 for CFBS. This performance level cannot be achieved by any of the reference methods, even when using the maximum number of available cameras $n = 9$, highlighting the clear performance and hardware efficiency gain of HFBS.
	
Although the BF solution remains the best-performing method for all $n$, HFBS performs comparably well for $n = 3,\dots,7$, while maintaining practical runtimes. Averaging over $n = 3,\dots,9$, HFBS achieves an average WCO of 237 compared to 334 for CFBS, corresponding to a relative reduction of 28.9\% in classification error. These results highlight that HFBS offers a near-optimal solution with both high accuracy and reduced hardware requirements, making it the most effective and practical method among the evaluated approaches.
	\begin{table}
		\caption{Average number of WCO of different approaches from literature for a fixed number of required cameras $n$.} 
	\centering
	\vspace{0.1cm}
	\begin{tabular}{l|p{0.34cm}p{0.34cm}p{0.34cm}p{0.34cm}p{0.34cm}p{0.34cm}p{0.34cm}}
	\backslashbox{Method}{$n$}& ~\textbf{3}~ & ~\textbf{4}~ & ~\textbf{5}~ & ~\textbf{6}~ & ~\textbf{7}~ & ~\textbf{8} & ~ \textbf{9}\\
	\hline
		
	Uniform \cite{UniformSearch} & 809 & 754 & 688 & 612 & 521 & 438 & 350 \\
	MOCR \cite{Hardenberg} & 637 & 519 & 431 & 402 & 358 & 335 & 320 \\ 
	FBS \cite{FBS} & 614 & 508 & 422 & 396 & 352 & 331 & 318\\
	CFBS \cite{CFBS} & 488 & 259 & 192 & 173 & 111 & 129 & 981\\
	HFBS (ours) & 144 & 124 & 118 & 89 & 73 & 129 & 981 \\
	BF (optimal) & 136 & 118 & 110 & 73 & 67 & 59 & 31 \\
\end{tabular}
\label{tab:fixnvaryingWCO}
\vspace{-0.4cm}
\end{table}

It can be concluded that, for classification tasks, our \mbox{HFBS} emerges as the most effective method. For cost-efficient \-set\-ups, it can determine the minimal number of cameras required to achieve a specified accuracy. For quality-focused applications, \mbox{HFBS} reliably delivers the highest classification precision for a fixed $n$, hence balancing efficiency and performance.

\begin{table}[t!]
\centering
\caption{Amount of filter combinations that have to be evaluated for the most effective solutions FBS, CFBS, HFBS and the optimal brute-force selection BF, as well as the respective processing times on a NVIDIA RTX A 4000 GPU.}
\vspace{0.1cm}
\begin{tabular}{l|c|c}
	Method & \# of evaluations &Runtime\\ 
	\hline
	Uniform search \cite{UniformSearch} & $> 1\times10^{13}$ & 3.06 min\\
	MOCR \cite{Hardenberg} & $> 8\times10^{13}$ & 24.46 min\\
	FBS \cite{FBS} & $> 8\times10^{12}$ & 110.09 s\\
	CFBS \cite{CFBS} & $> 6\times10^{7}$ & 12.41 s\\
	HFBS (ours) & $> 6\times10^{7}$ & 12.41 s\\
	BF (optimal) & $> 1\times10^{15}$ & 5.10 hrs\\
\end{tabular}
\label{tab:compCost}
\vspace{-0.4cm}
\end{table}

\vspace{-0.3cm}
\section{Conclusion}
\label{sec:Conclusion}
\vspace{-0.3cm}
This paper introduced a novel method to improve multispectral object classification performance using a hierarchical filter band selection. The HFBS performs two distinct classification steps to derive the material class and object label. A subsequent \-de\-cision tree refines these results, minimizing misclassifications and thus improving the accuracy. Further, HFBS allows optimization either for maximum classification accuracy given a fixed number of required cameras, or for the minimal number of required cameras necessary to achieve a predefined accuracy level. The proposed approach is particularly designed for applications where resource constraints, such as hardware limitations, need to be balanced against performance requirements.

%\balance

\bibliographystyle{IEEEbib}
\bibliography{refs.bib}

\pagebreak
\vfill
\enlargethispage{-0.9\baselineskip}

\end{document}